# PASSIVELY Q-SWITCHED YTTERBIUM AND CHROMIUM ALL-FIBRE LASER

Bernard Dussardier[(1)], Jérôme Maria[(1)], Pavel Peterka[(2)]
(1) Laboratoire de Physique de la Matière Condensée, Université de Nice Sophia-Antipolis, CNRS, UMR 6622, Parc Valrose, 06108 Nice Cedex 2, France
(2) Institute of Photonics and Electronics, Academy of Science of the Czech Republic, v.v.i., 18251 Prague, Czech Republic
bernard.dussardier@unice.fr

**Abstract:** A chromium-doped saturable absorber fibre stabilizes an otherwise spontaneously chaotic ytterbium-doped fibre laser. This original passively Q-switched all-fibre laser produces sustained and stable trains of smooth pulses at high repetition rate.

## 1. INTRODUCTION

The development of Q-switched fibre lasers is an alternative to bulk pulsed solid-state lasers operating in the ns- to µs-range for many applications such as material processing, medicine, aerial or space communications, remote sensing in free space or in fibre systems. Fibre lasers benefit from the high versatility and reliability of integrated, factory-aligned and fusion spliced fibre devices, relying on fiberized components produced at low cost and having high damage threshold. However Q-switched fibre lasers use externally driven intra-cavity bulk modulators, or bulk passive components such as a bulk polarizer [1], or a semi-conductor saturable absorber mirror [2]. These must be located in a free space section of the cavity, causing alignment and reliability problems.

The development of "all-fibre" passively Q-switched (PQS) fibre lasers is interesting because no external electronic driver is needed though some may need additional cheap and slow feedback control systems. Transition metal-doped bulk crystals have been used as saturable absorber (SA) within PQS fibre lasers [3,4], however alignment and optical damage were critical. We have for the first time proposed a core-pumped, low power, all-fibre PQS laser [5] using a newly developed chromium-doped fibre [6] as integrated SA (CrSA). Since, few PQS fiber lasers have implemented alternative SA ions like rare-earth (holmium [7], samarium [8]) or metals (bismuth [9]). However, some of these SA have drawbacks: low saturation intensity, long response time, absorption band narrower or wavelength-shifted relative to the gain medium. These characteristics necessitate additions to the laser cavity rendering the device more complex [7,9]. As for bulk SA, some transition metals from the 3d series, like $Cr^{4+}$ ions, are in principle the most attractive SA in silica, because of their high transition strength over broad wavelngth ranges [10] and short response time [11].

Other fibre laser systems produce self-pulsing in the sub-µs domain. Regular self-pulsing in erbium highly-doped fibres were demonstrated, based on energy transfers within the gain medium [12]. However this principle allows only a limited potential for engineering and optimization. More recently, double-clad ytterbium-doped fibre (DCYF) lasers have raised a great interest. However DCYF lasers have the particularity to randomly self-pulse in some cases. In particular, linear cavities with high loss and/or non uniform population inversion induce chaotic pulse generation [13]. The two main operating regimes are the sustained self pulsing (SSP, period longer than cavity round-trip time) and the self mode locking (SML, period equal to cavity round-trip time) [14].

In this paper, we take advantage of the self-pulsing in DCYF lasers and study the effect of a CrSA fibre on the laser dynamics. We show that the DCYF laser alone has a chaotic pulsed behaviour (both SSP and SML) whatever the pump level above threshold, whereas in presence of the Cr-doped fibre SA, relatively stable PQS pulses trains are produced.

## 2. EXPERIMENTALS
### 2.1 Fiber Characteristics

The DCYF and the single-mode CrSA fibre were prepared at LPMC by MCVD and solution doping technique. The DCYF has a D-shaped cladding (125 x 100 µm) for high pump absorption (3.8 dB/m at 975 nm). The diameter and NA of its single-mode core are 7 µm and 0.11, respectively. The CrSA was doped with $Cr^{4+}$ ions only, having a broadband absorption band across the visible and infrared wavelength range [6] and a composite SA relaxation behaviour [11]. Its core absorption is 3.5 dB/m at 1064 nm and mode-size adapted to the DCYF.

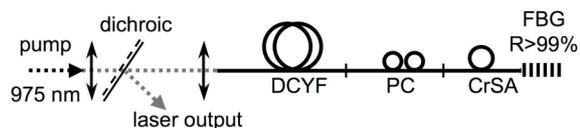

Fig. 1: Laser setup. See text.

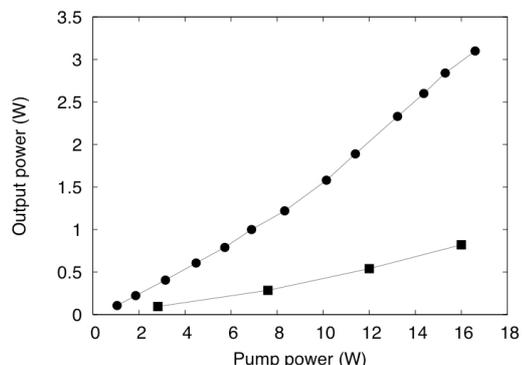

Fig. 2: Laser output vs incident pump power. ●: only DCYF, ■: DCYF + CrAF. Thresholds are 0.45 and 1.7 W, slope efficiencies 22% and 6%, maximum output power 3 W and 0.9 W, respectively

### 2.2 Fiber Laser Setups

Two laser setups were studied. The first used only the DCYF (length 1.2 m, pump absorption < 5 dB). The cavity was formed by a fibre Bragg grating (FBG) with high reflectivity (> 99,5 %) at 1064 nm (bandwidth 0.3 nm) and the perpendicular cleaved end of the DCYF (Fig. 1). The cavity was ~ 6.1 m long. The collimated pump beam was launched into the DCYF cladding using an aspherical lens. The polarization controller (PC) was used to partially control the laser stability. The FBG was limiting the number of longitudinal modes. Note that the pump coupling was not optimized, therefore pumping of the Yb ions populations was not homogeneous, providing favorable conditions for self pulsing (SSP). Also, this cavity was designed to self generate short pulses, therefore a short cavity length, a relatively high loss and high Yb ions inversion factor were chosen. As a consequence, the laser slope efficiency was not optimized. That also increased the stimulated Brillouin scattering threshold far above the achieved output power in this study.

The backward output beam was directed using a dichroic mirror toward a power-meter or a 1 GHz InGaAs detector connected to a fast oscilloscope or an optical spectrum analyser. In the second setup, 0.3 m of CrSA was spliced between the FBG and the PC. (Fig.1).

### 2.3 Results

The characteristics of both lasers are shown on Fig. 2. Its threshold and efficiency viz. the incident pump power were 0.45 W and 22 %, respectively. The DCYF laser had a chaotic behaviour spontaneously switching between CW and a Q-switched modes over the pumping range. A characteristic time trace of the output is shown on Fig. 3(a), typical of such cavity [3-5,7-9,11,13]. The few µs-long pulse envelops were strongly modulated by short pulses with the period of one cavity return-trip time (Fig. 3(b)). Here we observe a combination of SSP and SML. Using the PC to stabilize the output was quite inefficient, and at best produced a very noisy laser output. The chaotic behaviour was always dominant. Another short section of a commercially available Yb-heavily doped fibre, implemented in the same conditions, produced the same results.

By inserting a piece of CrSA fibre in the cavity, the characteristics of the laser changed drastically : first,

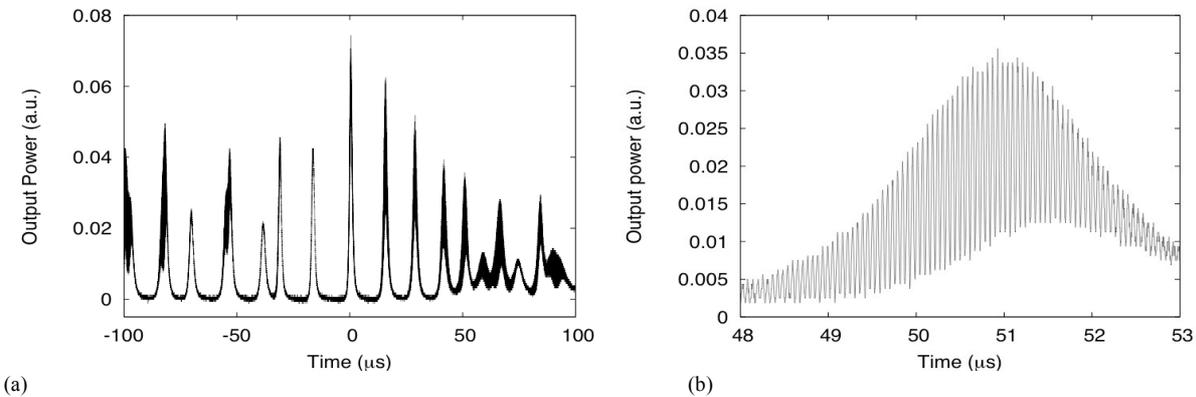

**Fig. 3:** (a) DCYF laser output, for 1W of pump (2.2 times threshold). (b) zoom on one typical pulse envelop.

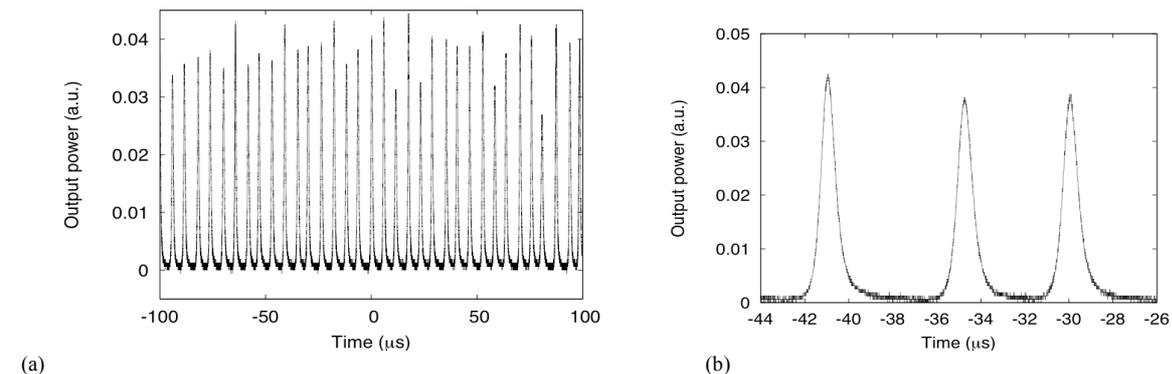

**Fig. 4:** (a) DCYF+CrSA laser output, for 7.6 W of pump (7.6 times threshold). (b) zoom on a typical pulses series.

the threshold of the laser increased and its efficiency diminished, as usually observed when inserting a SA in a cavity (Fig. 2). The CrSA fibre caused the threshold to increased upto 1.0 W, and the efficency to lower downto 6 %.

More importantly, the µs-long pulse trains were efficiently stabilized over the whole available pump range. On Fig. 4(a) is shown a typical output at 7.6 times the threshold. The peak RMS fluctuations were less that 10 % at 9.4 times the threshold. The higher the pump, the better the stability. Up to the maximum available pump (16 W), the stable passive Q-switching was still present ; no transition to CW operation was observed. The Q-switched envelopes were smoothed compared to the DCYF laser. No sub-modulation at the return trip period was visible anymore (Fig. 4(b)). The pulse duration and repetition rate versus incident pump power of the DCYF+CrSA laser are shown on Fig. 5, with a 400 ns minimum pulse width, a repetition rate as high as 350 kHz, a 2 µJ pulse energy and peak power equal to 6.4 W.

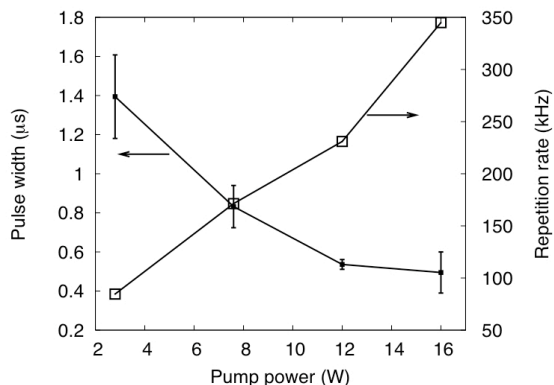

**Fig. 5:** Pulse duration and repetition rate of the passively Q-switch DCYF+CrSA laser.

### 3. DISCUSSION & CONCLUSION

The behaviour of the DCYF+CrSA laser is typical of a PQS mode of operation. It must be noted that this results were obtained with a set of cavity parameters causing self-pulsing or chaotic behaviour in the DCYF laser. Indeed, when CW or relatively stable output beam was set in the DCYF laser (by choosing parameters or setting the PC), then adding the CrSA provided only a quasi-CW regime. It is evident near the threshold of the DCYF+CrSA laser, when the pulse peak power is still too weak to induce the CrSA fibre to saturate : then the behaviour is fully chaotic. It regularly shifts toward a stable Q-switched regime when the pump increases. The capacity of our CrSA to stabilize SSP and cancel out SML in an Yb-doped fibre laser has never been reported, to the best of our knowledge.

In this particular cavity design, the shortest possible pulses were achieved. To further shorten them, a more compact cavity comprising only the active fibres (1.5 m) would produce sub-100-ns-pulses. Then the pulse peak power would be increased, causing a stronger CrSA saturation, and hence the pulse energy and the slope efficiency would be larger. In order to stay below the non-linear threshold, mode size enlargement will be required by introducing large-mode area fibre technology. MOPA configurations using a power DCYF amplifier could also be considered.

In conclusion, we have shown the potential of a CrSA fibre to produce stable passively Q-switched pulses trains out of a DCYF laser. This is to our knowledge, the first demonstration of such a laser with the active elements Yb and Cr. This laser architecture could be integrated and power upgraded, with great potentials for many applications.

### ACKNOWLEDGEMENT
"Fédération Doeblin" (CNRS, France); grant MSMT-Kontakt ME10119 (Czech Rep.); Czech-French 'Barrande' program 17360VA. LPMC is with GIS 'GRIFON' (CNRS, www.unice.fr/GIS). Visit us at http://lpmc.unice.fr/